\documentclass[twocolumn]{aastex6}
\usepackage{float}
\usepackage{graphicx}
\usepackage{amsmath}
\usepackage{natbib}
\usepackage{color}
\usepackage{verbatim}
\usepackage[utf8]{inputenc}
\usepackage{mathtools}
\usepackage{amssymb}
\usepackage{textcomp}

\pdfoutput=1

\DeclareUnicodeCharacter{2212}{\textminus}

\shortauthors{Smercina \textit{et al.}}

\accepted{to ApJL, 9 June 2017}
\submitted{Accepted to ApJL, 9 June 2017.}

\begin{document}
\title{d1005+68: A New Faint Dwarf Galaxy in the M81 Group}

\author{%
Adam Smercina\altaffilmark{1},
Eric F. Bell\altaffilmark{1}, 
Colin T. Slater\altaffilmark{2}, 
Paul A. Price\altaffilmark{3}, 
Jeremy Bailin\altaffilmark{4},
Antonela Monachesi\altaffilmark{5,6,7}
}

\altaffiltext{1}{Department of Astronomy, University of Michigan, 1085 S. University Avenue, Ann Arbor, MI 48109-1107, USA; asmerci@umich.edu}
\altaffiltext{2}{Astronomy Department, University of Washington, Box 351580, Seattle, WA 98195-1580, USA}
\altaffiltext{3}{Department of Astrophysical Sciences, Princeton University, Princeton, NJ 08544, USA}
\altaffiltext{4}{Department of Physics and Astronomy, University of Alabama, Box 870324, Tuscaloosa, AL 35487-0324, USA}
\altaffiltext{5}{MPA, Garching, Germany}
\altaffiltext{6}{Departamento de F\'isica y Astronom\'ia, Universidad de La Serena, Av. Juan Cisternas 1200 N, La Serena, Chile}
\altaffiltext{7}{Instituto de Investigaci\'on Multidisciplinar en Ciencia y Tecnolog\'ia, Universidad de La Serena, Ra\'ul Bitr\'an 1305, La Serena, Chile}

\begin{abstract}
We present the discovery of d1005+68, a new faint dwarf galaxy in the M81 Group, using observations taken with the Subaru Hyper Suprime-Cam. d1005+68's color-magnitude diagram is consistent with a distance of $3.98_{-0.43}^{+0.39}$~Mpc, establishing group membership. We derive an absolute \emph{V}-band magnitude, from stellar isochrone fitting, of $M_{V} = -7.94_{-0.50}^{+0.38}$, with a half-light radius of $r_{h} = 188_{-41}^{+39}$~pc. These place d1005+68 within the radius--luminosity locus of Local Group and M81 satellites and among the faintest confirmed satellites outside the Local Group. Assuming an age of 12 Gyr, d1005+68's red giant branch is best fit by an isochrone of [Fe/H] $= -1.90 \pm 0.24$. It has a projected separation from nearby M81 satellite BK5N of only 5 kpc. As this is well within BK5N's virial radius, we speculate that d1005+68 may be a satellite of BK5N. If confirmed, this would make d1005+68 one of the first detected satellites-of-a-satellite.  
\end{abstract}

\section{Introduction}
The past decade has seen an awakening in the field of dwarf galaxy discovery. Large photometric surveys such as the Sloan Digital Sky Survey (SDSS), the Panoramic Survey Telescope Rapid Response System (Pan-STARRS), and the Dark Energy Survey (DES) have permitted the discovery of $>$30 faint and ultrafaint dwarf galaxy (UFD) candidates in the Local Group \citep[e.g.,][]{belokurov2006,martin2013,drlica-wagner2016,homma2016}. These discoveries have informed the nearly two-decade-old ``missing satellites problem'' (hereafter MSP; \citealt{klypin1999}). This apparent tension between the low-end halo mass function slope, predicted by $\Lambda$CDM, and the considerably flatter slope of the Milky Way dwarf galaxy luminosity function is a sensitive probe of dark matter properties and galaxy formation in the lowest-mass dark matter halos \citep[e.g.,][]{maccio2010,brooks2013}. Yet, with improved understanding, new puzzles have emerged. An apparent dearth of luminous high-velocity subhalos -- the ``too big to fail'' problem (hereafter TBTF; \citealt{boylan-kolchin2011}) --- is an extension of MSP that is not alleviated by the discovery of UFDs (see \citealt{simon2007}, \citealt{maccio2010}, \citealt{font2011}, and \citealt{brooks2013} for discussion of possible solutions to MSP and TBTF). Furthermore, mounting evidence suggests that both the Milky Way's and M31's satellites form potentially planar structures \citep{pawlowski2013}. Though $\Lambda$CDM predicts anisotropic accretion due to infall along cosmic filaments (e.g., \citealt{li&helmi2008}), potentially resulting in planar satellite distributions \citep{sawala2016}, the thinness of the Local Group planes remains difficult to replicate. 

$\Lambda$CDM predicts that all galaxy halos host subhalos, the most massive of which will host luminous satellites. Consequently, many of the satellites around Milky Way--mass galaxies also likely possess, or possessed before infall, their own orbiting subhalos. These ``satellites-of-satellites'' are difficult to detect, owing to their intrinsic faintness. Recent work suggests that several of the Milky Way satellites nearest to the Magellanic Clouds may be satellites of the Clouds themselves \citep{drlica-wagner2016}, with possibly $>30$\% of Milky Way satellites originating around the Large Magellanic Cloud \citep[LMC;][]{jethwa2016}. 

It is clear that our understanding of dwarf galaxy populations in the $\Lambda$CDM paradigm is currently limited. A key hurdle is that our understanding of dwarf galaxy luminosity functions, spatial distributions, and properties is almost entirely confined to the Local Group. Characterization of satellite populations around other Local Group analogs is crucial if we are to obtain a complete description of low-mass galaxy formation. 

Propelled by the advent of wide-field imagers on large telescopes, discovery and characterization of faint `classical dwarfs' ($M_{V} < -10$) has become possible in nearby galaxy groups and clusters using large area (approaching 100 deg$^{2}$) diffuse light surveys \citep[e.g.,][]{chiboucas2009,mueller2015,munoz2015,ferrarese2016}. Observationally expensive, smaller area deep surveys of resolved stellar populations in nearby galaxy groups are bringing even fainter dwarf galaxies within reach \citep[e.g.,][]{sand2015,carlin2016,crnojevic2016,toloba2016}. 

In this Letter, we present the discovery of a faint dwarf spheroidal galaxy in the M81 group, d1005+68 (following the naming convention of \citealt{chiboucas2013}), detected as an overdensity of stars in observations taken with the Subaru Hyper Suprime-Cam. At $M_{V} = -7.9$~(see \S\,\ref{sec:gprop}), d1005+68 is one of the faintest confirmed galaxies discovered outside of the Local Group.

\begin{figure*}[t]
\centering
\leavevmode
\includegraphics[height=8.5cm,width={0.95\linewidth}]{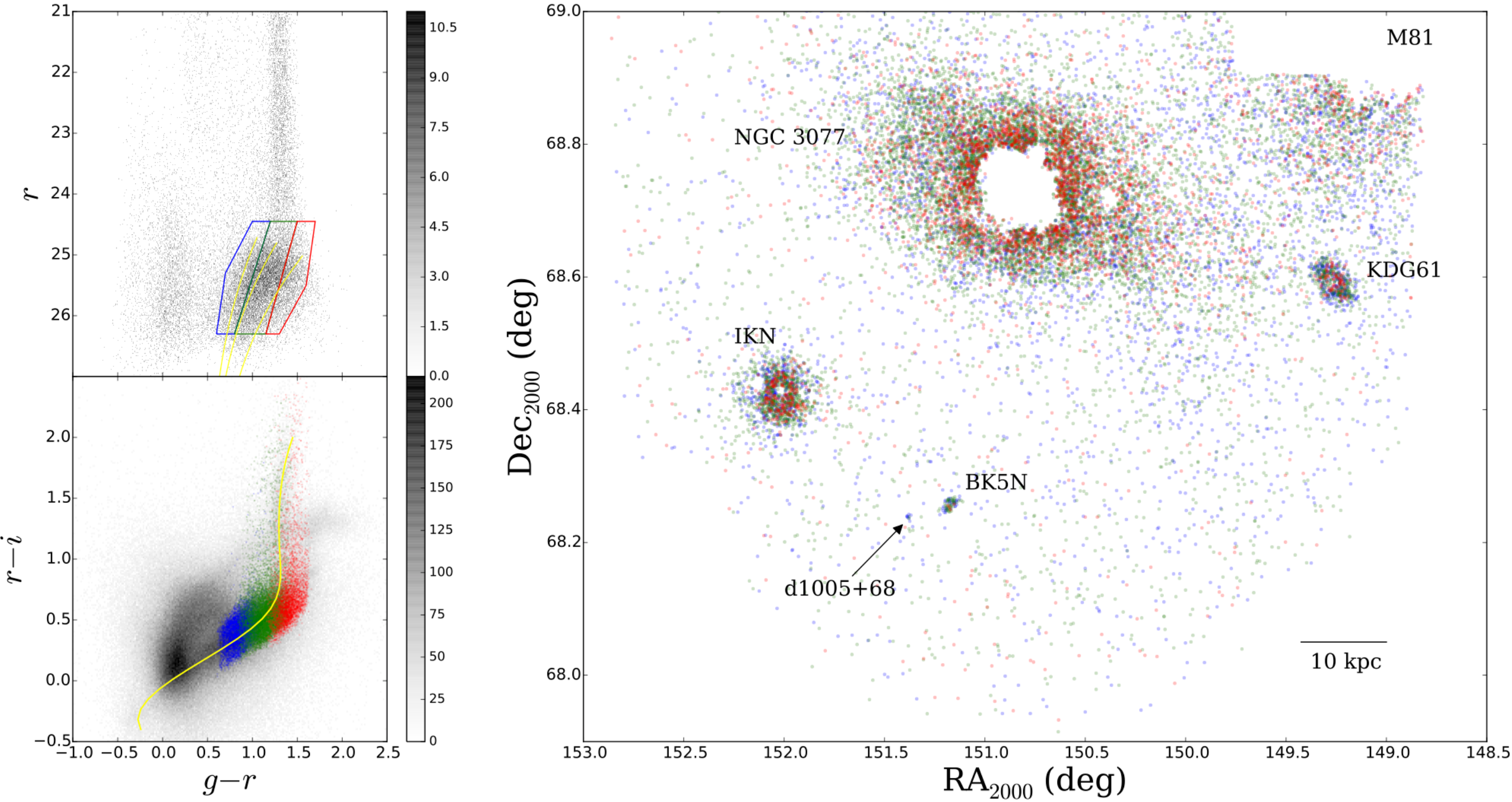} 
\caption{\uline{Top left}: the $g-r$ vs. $r$~CMD (de-reddened) of all stars (see \S\,\ref{sec:data}) in the Subaru field, separated into $\sim$0.01 mag bins. The RGB is encapsulated within the drawn polygon, which has been divided into three metallicity bins by eye, blue being the most metal-poor. The blue locus is likely a combination of young Helium burners and unresolved high-redshift background galaxies. The stripe at bright magnitudes is composed of Milky Way foreground stars. The yellow lines are 12 Gyr PARSEC isochrones \citep{bressan2012}, with [Fe/H] $= -2.1$~(left), [Fe/H] $= -1.7$~(center), and [Fe/H] $= -1.2$~(right), shown here for reference. \uline{Bottom left}: the $g-r$ vs. $r-i$ color-color diagram of photometrically identified sources. The stellar locus \citep{high2009} is shown as a yellow curve. RGB stars defined by our morphological, CMD, and stellar locus criteria (\S\,\ref{sec:data}) are shown as either blue, green, or red points, corresponding to their metallicity bin. The darkest region is the galaxy locus. \uline{Right}: a cutout of the map of M81's stellar halo in resolved RGB stars (Smercina et al. 2017, in preparation). The colors correspond to the metallicity bins defined on the CMD in the top left figure. The known galaxies in the field are labeled. d1005+68 is located at the bottom left of the map, indicated by a black arrow. It appears as a significant overdensity of blue (metal-poor) RGB stars, very near to the dwarf spheroidal, BK5N.}
\label{fig:rgb}
\end{figure*}

\begin{figure*}[t]
\centering
\leavevmode
\includegraphics[width={0.95\linewidth}]{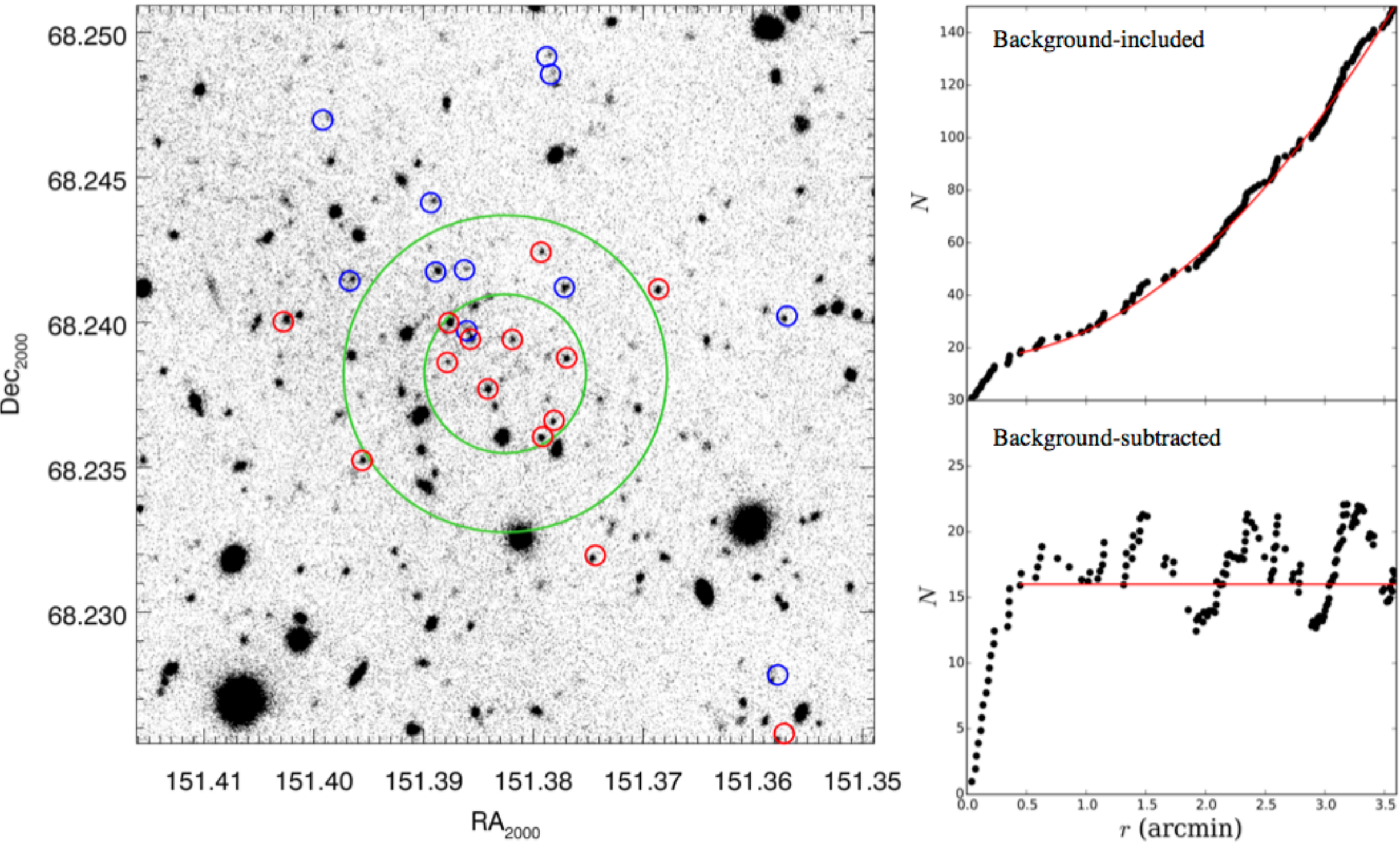}
\caption{\uline{Left}: the HSC $i$-band image of d1005+68. The concentric green circles correspond to apertures with 1 and 2$\times$~the derived half-light radius, centered on the estimated centroid. Member stars are encircled, with stars passing the 0\farcs84 size cut shown in red and those passing the broader 1\farcs34 size cut in blue (see \S\,\ref{sec:gprop}). \uline{Top right}: the curve of growth for d1005+68, using RGB stars defined by the 1\farcs34 size cut (the union of the blue and red stars). The red curve corresponds to an $N = \Sigma_{BG} r^{2}$~model of the background, using the derived Poisson mean with a 10\% correction ($\Sigma_{BG} \sim 3.3$~RGB stars arcmin$^{-2}$). \uline{Bottom right}: the background-subtracted curve of growth. The red line denotes the median value of $N - \Sigma_{BG}\pi r^2$, which we take as the number of member stars. The ``sawtooth'' nature of the radial profile is simply due to random over- and underdensities in the halo.
}
\label{fig:dc1}
\end{figure*}

\begin{flushleft}
\begin{deluxetable}{ll}
\tablecaption{\textnormal{d1005+68 Parameters}\label{tab:dc1}}
\tablecolumns{2}
\setlength{\tabcolsep}{10pt}
\setlength{\extrarowheight}{0.7pt}
\tablewidth{0.95\linewidth}
\tabletypesize{\small}
\tablehead{%
\colhead{Parameter} &
\colhead{Value} \\
}
\startdata
$\alpha$ (J2000) & $10^{h} 05^{m}31\fs 82 \pm 1\fs 1$ \\
$\delta$ (J2000) & $+68^{\circ} 14\arcmin 19\farcs 56 \pm 5\farcs 95$ \\
$D_{\textrm{TRGB}}$ & $3.98^{+0.39}_{-0.43}$~Mpc \\
$M_{V}~^{a}$ & $-7.94^{+0.38}_{-0.50}$ \\
$r_{h}$ & $9\farcs 7 \pm 2\farcs 0$\\
$r_{h}$ & $188^{+39}_{-41}$~pc \\
$\log_{10} (M_{*}/M_{\odot})~^{b}$ & $5.40^{+0.22}_{-0.16}$ \\
$[$Fe/H$]~^{c}$ & $-1.90 \pm 0.24$ \\
\enddata
\tablecomments{$^{a}$~Isochrone fitting, assuming $D_{\textrm{TRGB}}$. $^{b}$~Current stellar mass, assuming 40\% mass loss. $^{c}$~Metallicity of best-fit isochrone, assuming $[\alpha$/Fe] = 0.25.}
\end{deluxetable}
\end{flushleft}

\section{Detection}
\label{sec:data}

We use observations taken with the Subaru Hyper Suprime-Cam (HSC; \citealt{miyazaki2012}) through NOAO
Gemini-Subaru exchange time (PI: Bell, 2015A-0281). The observations consist of two pointings for a survey footprint area of $\sim 3.5$~deg$^{2}$, in three filters: $g$, $r$, and $i$, with $\sim 3600$\ s per filter per pointing. The data were reduced using the HSC pipeline (Bosch et al. 2017, in preparation), which was developed from the LSST Pipeline \citep{axelrod2010}. The data were calibrated using photometry and astrometry from Pan-STARRS1 \citep{magnier2013}. An aggressive background subtraction using a 32 pixel region for determining the background was used. Objects are detected in $i$\ band and forced photometry is performed in $g$~and $r$. The average FWHM in M81 Field 2 (in which d1005+68 was discovered) is $\sim$0\farcs7 in all bands, giving limiting 5$\sigma$~point-source magnitudes of $g \sim 27$, $r \sim 26.5$, and $i \sim 26$. All magnitudes use the SDSS photometric system, corrected for foreground Galactic extinction using the \citet{sfd-extinction} maps as calibrated by \citet{schlafly2011}. 

As the dwarf galaxies of interest are low surface brightness and possess little diffuse emission, we detect dwarf candidates by resolving them into individual stars. At the distance of M81 (3.6 Mpc; \citealt{radburn-smith2011}), only stars in the top $\sim$25\%, or tip of the RGB (TRGB), are visible. TRGB stars are relatively numerous, and as they trace the old stellar population of galaxies, their number can be scaled to a total luminosity with modest uncertainty \citep{harmsen17}.

At our survey depths, contaminants -- high-redshift background galaxies -- dominate. The majority of these galaxy contaminants must be removed in order to reach the surface brightness sensitivity necessary to detect faint dwarf satellites ($\mu_{V} \lesssim 28$~mag arcsec$^{2}$). We reject galaxies using a combined morphology and color cut; such a process sacrifices completeness in order to dramatically suppress contamination (this will be revisited in \S\,\ref{sec:gprop}). To be defined as a star, a source must satisfy two criteria: (1) FWHM $\leqslant0.6\arcsec$~across all three bands (we will consider less stringent cuts later), and (2) consistent with the $g-r$ vs. $r-i$ stellar locus within $\sigma_{g-r}$~(the photometric uncertainty) + 0.2 mag (intrinsic scatter; \citealt{high2009}). Next, we locate stars on the RGB from the $g-r$ vs. $r$ color-magnitude diagram (CMD) and divide them into three metallicity bins using simple polygonal boundaries (see Figure \ref{fig:rgb}). 

d1005+68 stands out as a significant overdensity of metal-poor stars in the sparse, metallicity-binned RGB star map of M81's stellar halo (Figure \ref{fig:rgb}), with nine RGB stars visible in a $1\arcmin \times 1\arcmin$ region centered on d1005+68. To quantify the prominence of this overdensity against the surrounding diffuse stellar halo, we extract 500 $1\arcmin \times 1\arcmin$~(independent) regions from a 0.14 deg$^{2}$ region south of d1005+68, away from the stellar debris associated with the tidal disruption of NGC 3077. We compute the discrete probability distribution of the number of RGB stars returned in each region and fit it to a Poisson distribution, $p(N|\lambda)$. From the best-fit Poisson distribution, we take a mean background of $\lambda = 0.38 \pm 0.03$~RGB stars arcmin$^{-2}$. Integrating over the best-fit distribution, and correcting for the number of independent 1 arcmin$^{-2}$~regions ($10^{4}$) in the target footprint, we obtain a cumulative probability of drawing nine RGB stars arcmin$^{-2}$~of $4.2\times 10^{-6} \pm 3.5\times 10^{-6}$. Placed into terms of standard error, this is a $4.5-5\sigma$ detection. Thus, we expect to detect 0.01 such random overdensities in our target footprint.~In the following section, we discuss the derivation of d1005+68's properties, which are summarized in Table \ref{tab:dc1}. Its position relative to other M81 Group members is shown in the map of M81's stellar halo in Figure \ref{fig:rgb}. 
\begin{figure*}
\centering
\leavevmode
\includegraphics[width={0.95\linewidth}]{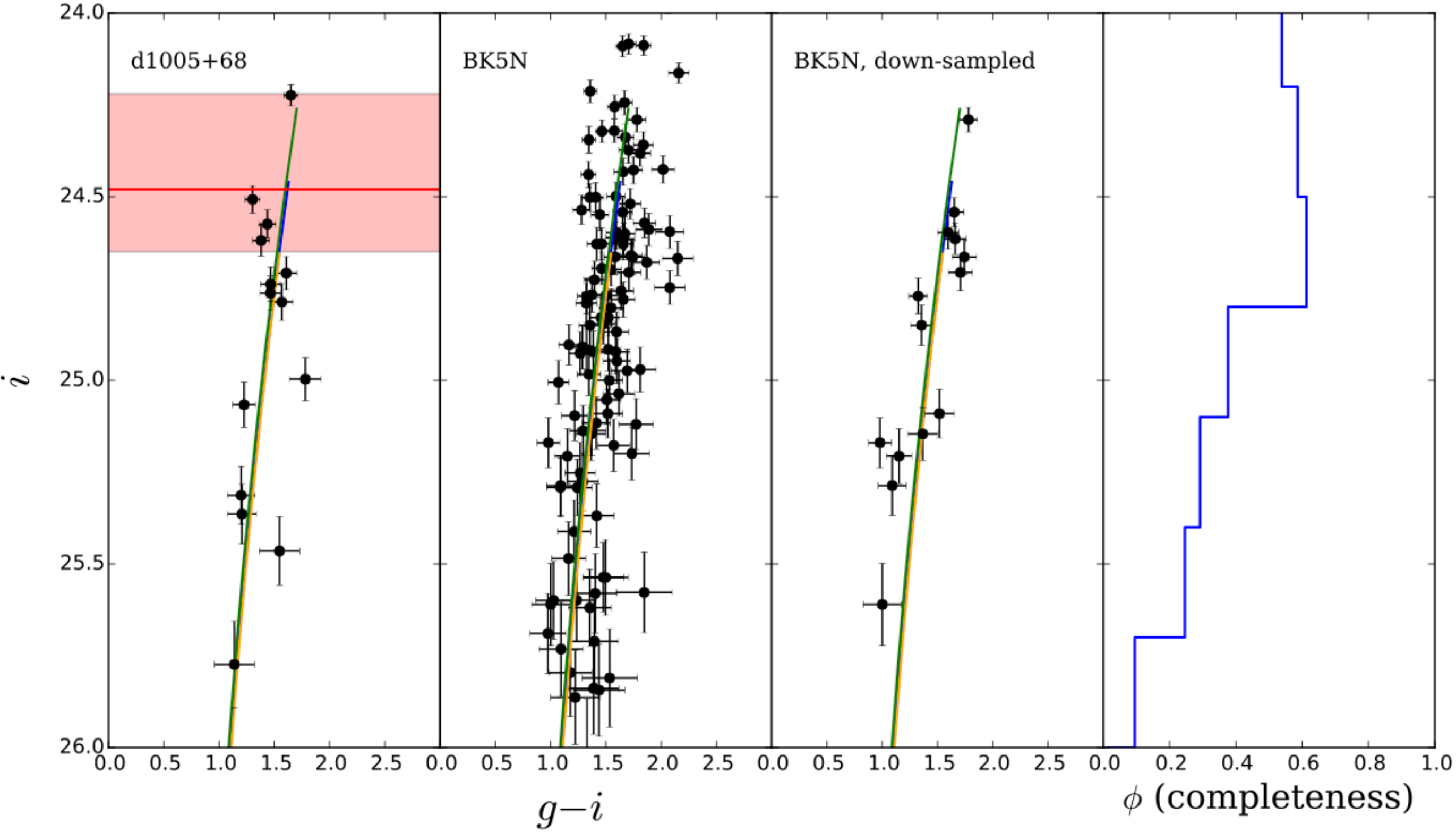}
\caption{\uline{Left}: the color-magnitude diagram of d1005+68. Stars shown are identified with the 0\farcs84 size cut (see \S\,\ref{sec:gprop}), extending to $\sim$0\farcm5 or $\sim$3$r_{h}$. The TRGB is shown as a red line, with the 90\% confidence shown as the red shaded region. The three blue curves on each diagram correspond to the best-fit 12 Gyr isochrones at each distance bound, with respective metallicities (from left to right) of [Fe/H] $= -1.76$~(green), $-1.90$~(blue), and $-2.02$~(orange). \uline{Center left}: The CMD of BK5N in RGB stars, with $\sim 100$\ detected RGB stars. \uline{Center right}: BK5N's CMD, randomly down-sampled to match the number of member stars in d1005+68. \uline{Right}: the $i$-band completeness function, $\phi$.
}
\label{fig:cmd_sat}
\end{figure*}
In Figure \ref{fig:dc1} we show the $i$-band image of d1005+68 with detected RGB stars encircled, as well as the curve of growth.

\section{Properties}
\label{sec:gprop}
The $g - i$ vs. $i$~CMD of probable member stars of d1005+68 are shown in Figure \ref{fig:cmd_sat}. We define membership based on the shape of the curve of growth (Figure \ref{fig:dc1}, bottom right panel), where the background-subtracted profile asymptotes to a $\sim$constant value. In contrast to the stringent cut used for detection of the dwarf, we use broader criteria for membership determination and the derivation of the dwarf's properties. At low signal-to-noise, the measured sizes of objects is subject to significant scatter, causing tight tolerances on size to reject many true stars. Consequently, the stars shown on the CMD were chosen using the same color constraint as for detection, but with a looser size constraint -- FWHM in $x$~and $y$~$\leqslant 0\farcs84$. Also shown in Figure \ref{fig:cmd_sat} are CMDs of nearby (in projection) dwarf galaxy BK5N -- both full and randomly down-sampled to the number of observed stars in d1005+68. 

The centroid, half-light radius, and number of member stars (and therefore luminosity) are the averages of a range of values estimated by varying the size cut between 0\farcs6 and 1\farcs34, the number of stars used to define the position of the center (relative to the optical center) between 5 and 12, and the Poisson background value (see \S\,\ref{sec:data}). For each iteration, the number of member stars are determined using the turnover of the background-subtracted curve of growth, from which the half-light radius is also derived. The mean values of the centroid and half-light radius can be found in Table \ref{tab:dc1}, along with the standard deviations of the various iterations. 

The TRGB can be used as a robust distance estimator, due to its near-constant luminosity ($M_{I} = -4.04$~in the Johnson-Cousins system) at low metallicities \citep{bellazzini2001}. The TRGB for d1005+68's CMD (see Figure \ref{fig:cmd_sat}) was calculated as in \cite{monachesi2016}, but also includes the completeness in the model luminosity function (LF), $\phi$~(see below), as in \cite{makarov2006}:
\begin{equation}
\phi(m | \mathbf{x}) = \int \psi(m^{\prime} | \mathbf{x}) \, e(m | m^{\prime}) \, \rho(m^{\prime}) \, dm^{\prime}
\end{equation}
where $\psi$ is the true LF, $e$ is the Gaussian error kernel, $\rho$ is the completeness, and $\mathbf{x}$ is the vector of model parameters that we fit. See Appendix C of \cite{monachesi2016} for details. The completeness was tabulated in 0.3 mag $i$-band bins using the area in common with GHOSTS and smoothed with a three-bin boxcar (the smoothing has no effect on the derived TRGB magnitudes). We find a TRGB of $i_{TRGB} = 24.48_{-0.26}^{+0.17}$. Using a SDSS ``Lupton prescription,'' \footnote{{\href{https://www.sdss3.org/dr8/algorithms/sdssUBVRITransform.php}{https://www.sdss3.org/dr8/algorithms/sdssUBVRITransform.php}}}, in the JC system, this corresponds to $I_{TRGB} = 23.96_{-0.25}^{+0.20}$, or a distance modulus of $m - M = 28.00_{-0.25}^{+0.20}$. Thus, we derive a distance to d1005+68 of $3.98_{-0.43}^{+0.39}$~Mpc. 

\begin{figure*}[t]
\centering
\leavevmode
\includegraphics[width={0.95\linewidth}]{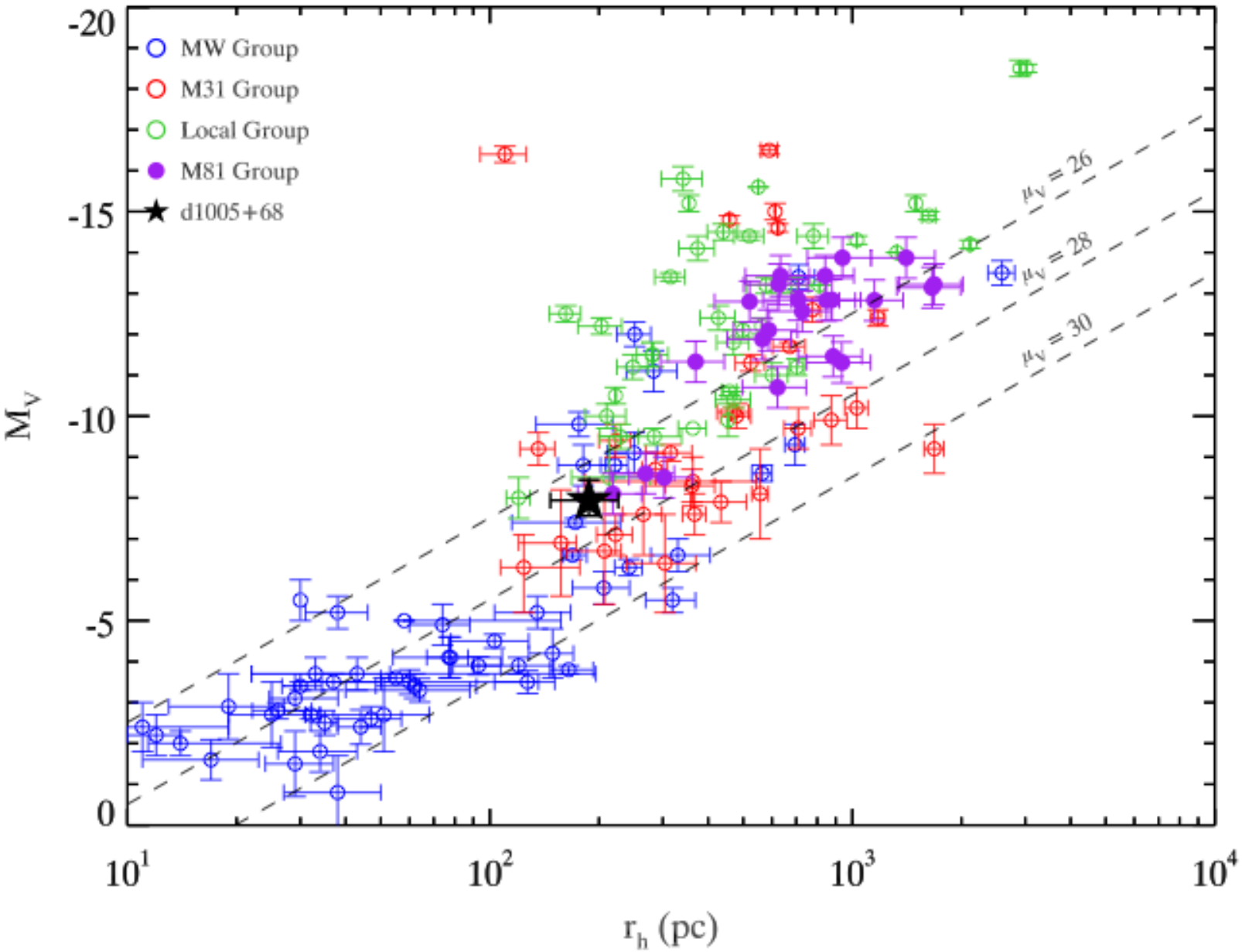}
\caption{Half-light radius--luminosity diagram for Milky Way, M31, Local Group, and M81 Group satellites. Milky Way satellites are shown as blue circles, M31 satellites as red circles, general Local Group members (outside the virial radius of MW or M31) as green circles, and M81 members as filled purple circles. Local Group data are compiled from the catalog of \cite{mcconnachie2012}, from the recent slew of Dark Energy Survey \citep{bechtol2015,drlica-wagner2015,koposov2015} and Pan-STARRS \citep{laevens2015a} discoveries, and from other isolated discoveries \citep{belokurov2014,kim2015,homma2016}. M81 Group data are compiled from \cite{karachentsev2000}, \cite{lianou2010}, and \cite{chiboucas2013}. In the absence of $M_{V}$\ and $r_{h}$\ uncertainties in the literature, typical Local Group uncertainties of 20\% have been adopted for M81 members. d1005+68 is shown as a black star. Lines of constant surface brightness are shown for reference. Our derived $r_{h}$\ and $M_{V}$\ for d1005+68 place it well within the locus of Local Group satellites, while it is one of the faintest members of the M81 Group.
}
\label{fig:rlum}
\end{figure*}

d1005+68's luminosity was estimated using the number of stars visible to a certain $i$-band ``depth" below the TRGB. To convert the number of observed stars to a total number of stars above this $i$-band limit, we use the GHOSTS fields for M81 \citep{radburn-smith2011} to compute the stellar completeness in the Subaru field, as a function of $i$-band magnitude, for our three size cuts (0\farcs6, 0\farcs84, 1\farcs34). For all three size cuts, we estimate a total number of $32 \pm 6$~RGB stars to a depth of $\sim$1.2 mag below the TRGB, and $25 \pm 4$~to a depth of $\sim$1.1 mag below the TRGB. We then randomly sample our best-fit isochrone in that magnitude range given a \citet{chabrier2003} stellar initial mass function (IMF). We record the resulting number of RGB stars drawn at each stellar mass and compute a probability distribution of drawing the observed number of stars at each mass, at the given RGB depth. We obtain a most probable initial mass of $\log_{10}(M_{*}/M_{\odot}) = 5.62$, which, after the standard 40\% mass-loss correction \citep{bruzual&charlot2003}, corresponds to a current stellar mass of $\log_{10}(M_{*}/M_{\odot}) = 5.40$~or $M_{*} = 2.5\times 10^{5} M_{\odot}$. We then convert the stellar mass distribution to a $V$-band luminosity, while randomly varying the number of stars in each isochrone, at a fixed stellar mass. Accounting for the variance in the different depths considered, as well as sampling variance along the IMF, we obtain a $V$-band luminosity of $M_{V} = -7.94_{-0.50}^{+0.38}$. The primary uncertainties on this estimate come from our TRGB distance range and the width of the best-fit stellar mass distribution. 

To estimate the metallicity, we fit a suite of PARSEC stellar isochrone models \citep{bressan2012} with a fixed 12 Gyr age, from $Z = 0.0001 - 0.001$. The best-fit isochrone, for the $g-i$~vs. $i$~CMD, corresponds to a metallicity of $Z = 0.0004$. Assuming [$\alpha$/Fe] $= 0.25$, this corresponds to [Fe/H] $= -1.90$. For each iteration in the centroid calculation (above) we draw 10 bootstrap samples and compute the best-fit isochrone for each case. We then combine the standard deviation of the resulting distribution with the TRGB distance uncertainties. We obtain a final metallicity estimate of [Fe/H] $= -1.90 \pm 0.24$. 

d1005+68 has a projected separation from M81 of 1\fdg22, or, using the distance to M81, $76.4$~kpc. Using the adopted TRGB distance to d1005+68 of $3.98_{-0.43}^{+0.39}$~Mpc, this corresponds to a large range in possible 3D distances. The projected physical separation between d1005+68 and the nearby (on the sky) dwarf spheroidal BK5N is only $\sim 5$~kpc at the distance of BK5N (3.78 Mpc; \citealt{karachentsev2000}). Assuming a stellar mass for BK5N of $\sim 10^7 M_{\odot}$~($M_{V} = -11.33$; \citealt{caldwell1998}) and extrapolating from the stellar mass--halo mass relation of \cite{behroozi2013}, the virial radius of BK5N is likely $\sim 40$~kpc. Therefore, were d1005+68 at a similar distance as BK5N, it would be well within BK5N's virial radius. In support of this, the CMD of d1005+68 is well approximated by a random sampling of BK5N's CMD, as in Figure \ref{fig:cmd_sat}. However, the 3D separation could be much higher when factoring in the uncertainty in d1005+68's TRGB distance. 

\section{Discussion and Closing Remarks}
\label{sec:discuss}

In this Letter, we presented a new faint dwarf galaxy, d1005+68, with properties consistent with being a satellite of the M81 Group.~It was detected as a $5\sigma$~overdensity in our 3.5 deg$^{2}$~Subaru Hyper Suprime-Cam survey of M81's resolved stellar halo. We find that the CMD is best fit by an isochrone of age 12 Gyr and metallicity [Fe/H] = $-1.90 \pm 0.24$. d1005+68 has projected physical distances from M81, NGC 3077, and BK5N of $\sim$76 kpc, 40 kpc, and 5 kpc, respectively. The estimated heliocentric TRGB distance of $3.98_{-0.43}^{+0.39}$~Mpc provides strong evidence for group membership; however, the high uncertainties prohibit accurate estimates of 3D separation from other group members. Its current stellar mass, determined from isochrone fitting, is $M_{*} = 2.5_{-0.8}^{+1.7} \times 10^{5} M_{\odot}$, corresponding to an absolute $V$-band magnitude of $M_{V} = -7.94_{-0.50}^{+0.38}$. 

Figure \ref{fig:rlum} shows d1005+68 in context of Local Group and M81 Group members. d1005+68 is among the faintest confirmed galaxies discovered outside of the Local Group -- similar in brightness to M81 group member d0944+69 (\citealt{chiboucas2013}; $M_{V} = -8.05$ with no claimed uncertainty), NGC 2403 member MADCASH J074238+652501-dw (\citealt{carlin2016}; $M_{V} = -7.7 \pm 0.7$), Centaurus group member Dw5 (\citealt{crnojevic2016}; $M_{V} = -7.2 \pm 1.0$), and Fornax cluster member Fornax UFD1 (\citealt{lee2017}; $M_{V} = -7.6 \pm 0.2$) -- and probes the very faintest end of the known M81 satellite luminosity function. 

The projected separation between d1005+68 and BK5N of 5 kpc is well within the estimated virial radius of BK5N ($\sim 40$~kpc). With our highly uncertain TRGB distance (due to scarcity of stars) and the similarity between the two CMDs (Figure \ref{fig:cmd_sat}), this introduces the \textit{possibility} that d1005+68 is a satellite of BK5N. If confirmed (via more accurate distance estimates and line of sight velocity information), this would make it the first satellite-of-a-satellite discovered outside of the Local Group. 

\acknowledgments
We thank the anonymous referee for helpful comments that substantially improved this Letter.

This material is based upon work supported by the National Science Foundation Graduate Research Fellowship Program under grant No.~DGE 1256260. Any opinions, findings, and conclusions or recommendations expressed in this material are those of the author(s) and do not necessarily reflect the views of the National Science Foundation. 

Based on observations obtained at the Subaru Observatory, which is operated by the National Astronomical Observatory of Japan, via the Gemini/Subaru Time Exchange Program. We thank the Subaru support staff -- particularly Akito Tajitsu, Tsuyoshi Terai, Dan Birchall, and Fumiaki Nakata -- for invaluable help preparing and carrying out the observing run. 

The Pan-STARRS1 Surveys (PS1) have been made possible through contributions of the Institute for Astronomy, the University of Hawaii, the Pan-STARRS Project Office, the Max-Planck Society and its participating institutes, the Max Planck Institute for Astronomy, Heidelberg and the Max Planck Institute for Extraterrestrial Physics, Garching, The Johns Hopkins University, Durham University, the University of Edinburgh, Queen's University Belfast, the Harvard-Smithsonian Center for Astrophysics, the Las Cumbres Observatory Global Telescope Network Incorporated, the National Central University of Taiwan, the Space Telescope Science Institute, the National Aeronautics and Space Administration under grant No. NNX08AR22G issued through the Planetary Science Division of the NASA Science Mission Directorate, the National Science Foundation under grant No. AST-1238877, the University of Maryland, and Eotvos Lorand University (ELTE).

\end{document}